\documentstyle[12pt]{article}

\newcommand{\eq}{\begin{equation}}
\newcommand{\en}{\end{equation}}
\newcommand{\bea}{\begin{eqnarray}}
\newcommand{\eea}{\end{eqnarray}}
\newcommand{\spz}{\hspace{0.7cm}}

\newcommand{\virg}{\spz,\spz}

\newcommand{\cG}{{\cal G}}
\newcommand{\cL}{{\cal L}}
\newcommand{\D}{\Delta}

\newcommand{\NP}[1]{Nucl.\ Phys.\ {\bf #1}}
\newcommand{\PL}[1]{Phys.\ Lett.\ {\bf #1}}

\newcommand{\PRL}[1]{Phys.\ Rev.\ Lett.\ {\bf #1}}
\newcommand{\MPL}[1]{Mod.\ Phys.\ Lett.\ {\bf #1}}
\newcommand{\IJMP}[1]{Int.\ J.\ Mod.\ Phys.\ {\bf #1}}

\begin{document}
\sloppy
\renewcommand{\thefootnote}{\fnsymbol{footnote}}

\newpage
\setcounter{page}{0}

\vskip 1cm
\begin{center}
{\bf THERMODYNAMIC BETHE ANSATZ FOR $\cG_k \otimes \cG_l / \cG_{k+l}$ COSET
MODELS PERTURBED BY THEIR $\phi_{1,1,Adj}$ OPERATOR}\\
\vskip 1.8cm
{\large F.\ Ravanini\footnote{Permanent address after March 1, 1992: INFN -
                              Sez. di Bologna, Via Irnerio 46, I-40126 BOLOGNA
                              - ITALY}}\\
\vskip .7cm
{\em Service de Physique Th{\'e}orique, C.E.A. - Saclay \footnote{
     Laboratoire de la Direction des Sciences de la Mati{\`e}re du
     Commissariat \`a l'Energie Atomique} \\
     Orme des Merisiers, F-91190 Gif-sur-Yvette, France\\
     and\\
     I.N.F.N. - Sez. di Bologna, Italy}
\end{center}
\vskip 1cm

\renewcommand{\thefootnote}{\arabic{footnote}}
\setcounter{footnote}{0}

\begin{abstract}
We propose a Thermodynamic Bethe Ansatz (TBA) for $\cG_k\otimes\cG_l
/\cG_{k+l}$ conformal coset models ($\cG$ any simply-laced Lie algebra)
perturbed by their operator $\phi_{1,1,Adj}$. An interesting adjacency
structure appears and can be depicted in a sort of ``product'' of Dynkin
diagrams of $\cG$ and $A_{k+l-1}$. UV and IR limits are computed and reproduce
the expected values for the central charges. For $k\to\infty$, $l$ fixed we
obtain the TBA of the $\cG_l$ WZW model perturbed by $J_a\bar{J}_a$, and for
$k,l\to \infty$, $k-l$ fixed, that of Principal Chiral model with WZ term at
level $k-l$.
\end{abstract}
\vskip .3cm
\begin{flushright}
Saclay preprint SPhT/92-011\\
January 1992
\end{flushright}
\vskip .3cm
Submitted for publication to Phys.Lett.B
\newpage

Many two dimensional Quantum Field Theories obtained by deformation of
a Conformal Field Theory (CFT) by one of its relevant operators show an
infinite number of conserved currents that guarantee integrability.
There have been various proposals to identify the factorizable
S-matrix of such a theory, by use of local~\cite{zam1}
or non-local~\cite{zam2,BeLec} conserved currents.

Recently Al.Zamolodchikov~\cite{Al1} proposed the use of {\it Thermodynamic
Bethe Ansatz} (TBA) techniques to analyze the ultraviolet (UV) and infrared
(IR) behaviours of a theory with given S-matrix
and to follow the RG flows of various
operators. Comparing the results with those expected from perturbed CFT, one
can provide a non-trivial check for the S-matrices conjectured via the methods
of~\cite{zam1,zam2,BeLec}.
For the case of {\it diagonal} S-matrices, the TBA system of
coupled non-linear integral equations determining the thermodynamics of
the relativistic scattering theory
are relatively simple to obtain and for large sets of such S-matrices they
have been explored~\cite{km}, the central charges, the scaling dimensions of
the perturbing operator and some other data of the UV theory being reproduced
as expected.

TBA equations are much more difficult to derive in the case of {\em
non-diagonal} S-matrices, where in principle one needs to diagonalize the
so called {\em color transfer
matrix}, via higher level Bethe ansatz, a formidable problem in many cases.
Nevertheless, TBA equations have been proposed for minimal models perturbed
by its least relevant operator $\phi_{13}$, for negative~\cite{ItMo,Al3}
values of the coupling. These theories have an S-matrix~\cite{BeLec}
describing the scattering of kinks, all of the same mass $M$, separating
vacua of different colors. Moreover, one can modify this set of TBA equations,
in order to describe the same model with positive coupling,
which consists in a massless theory flowing to the nearest lower model in the
stair of minimal CFT's~\cite{Al4}. Finally, another modification of the same
TBA allows to describe flows between generalizations of minimal models, namely
coset models of the kind $SU(2)_k \otimes SU(2)_l/ SU(2)_{k+l}$~\cite{Al5}
(that we shall denote for short $[A_1]^{k,l}$ in the following).
In~\cite{Al3,Al4,Al5} the
TBA equations are first conjectured on reasonable physical grounds, then
from them one can extract non trivial information that can be compared with
conformal perturbation theory to confirm that the proposed TBA does indeed
describe the expected CFT, deformed by the chosen operator. Once a TBA is
(out of reasonable doubt) attached to a certain deformed UV CFT, one can use
it to follow the Renormalization Group (RG) flow of the theory. In particular,
the IR behaviour can be recovered as the TBA equations are exactly solvable
in this limit. By use of this approach, Al.Zamolodchikov was able to check
explicitly some previously conjectured features of two-dimensional RG flows,
as e.g. the fact that the $[A_1]^{k,l}$ models perturbed by their least
relevant operator for positive coupling flow to the $[A_1]^{k-l,l}$ ones.
Even more surprisingly, some completely new RG flows connecting non trivial
UV and IR points have been discovered, e.g. V.Fateev and Al.Zamolodchikov
give evidence of new flows from
$Z_{k+1}$ parafermionic models to $[A_1]^{k,1}$ minimal models~\cite{FatAl}.

An interesting question is that
of trying to generalize these results to the large class of $\cG_k \otimes
\cG_l/\cG_{k+l}$ coset CFT's, i.e. to $[\cG]^{k,l}$ in the previous notation,
where $\cG$ is {\em any} Lie algebra~\cite{chr-rav}.
Unfortunately, for $\cG$ non-simply laced, the situation on the knowledge of
S-matrices is still confused, therefore in the following we shall restrict
to the class of simply-laced Lie algebras $A,D,E$.

Perturbing these models by their scalar relevant operator $\phi_{1,1,Adj}$,
\footnote{This operator has conformal dimension $\D=\frac{k+l}{k+l+g}$, $g$
being the dual Coxeter number of $\cG$. The 3 indices label $\cG_k$, $\cG_l$
and $\cG_{k+l}$ integrable representations respectively.}
in the present paper we consider the models $[\cG]^{k,l}_{\pm}$
defined by the following action
\eq
{\cal A}_{[\cG]^{k,l}_{\pm}} =
{\cal A}_{[\cG]^{k,l}} \pm \lambda \int d^2z \phi_{1,1,Adj} \virg \lambda
\geq 0
\en
In analogy with the known $\cG=A_1$ case, we expect that the model
$[\cG]^{k,l}_-$ will
be dominated by massive kinks connecting the different vacua of the theory,
while the model $[\cG]^{k,l}_+$ will define a massless theory flowing to
an IR limit identified with the $[\cG]^{k-l,l}$ CFT.

As a startpoint of our analysis, we consider the models $[\cG]^{k,1}_-$. The
integrability of these models has been discussed in~\cite{FatZam} and kink
S-matrices have been proposed in~\cite{DevFat,AhBeLec}.
They are non-diagonal, with
the exception of $k=1$, which corresponds to the case studied in~\cite{km}.
When a non-diagonal S-matrix appears, one has to diagonalize the color
transfer matrix. This leads to
Bethe equations, that can be interpreted as periodicity conditions on a system
of fictitious particles carrying no energy, usually called {\em magnons},
moving and scattering in a ``frame lattice'' of physical particles.
(see e.g.~\cite{houches} and refs. therein). To find the details of the
magnonic structure one should solve the Bethe equations, which is notably a
very difficult task. In the case we are examining, the diagonalization of
the color transfer matrix is equivalent to that for the transfer matrix of
an IRF model~\cite{jmo} at criticality. This diagonalization has been carried
out by
Bazhanov and Reshetikhin~\cite{BaRe}, who also studied the thermodynamic of
the corresponding one-dimensional model and of its scaling limit.
To be precise, in the solution
proposed in~\cite{BaRe} there are some assumptions: first of all the assumption
that the solution organize themselves in the thermodynamic limit as a set of
{\em strings}. The string hypotesis is not always true (see e.g.~\cite{AlbMcc})
but it is quite reasonable here. Moreover, by analogy with known cases,
they assume that some of the strings
have densities that vanish in the thermodynamic limit. Finally, all the
machinery is explicit for $A_N$ algebras only, and is then {\em conjectured}
for $D_N$ and $E_{6,7,8}$. Nevertheless, we shall see that these assumptions
lead to reasoanble results.

The TBA for our scattering theory that can be
deduced from~\cite{BaRe} can be presented in many ways. In this paper we write
two equivalent forms of it. The first one is perhaps more suitable for
practical calculations, while the second shows a high degree of universality
(in the spirit of ref.~\cite{Al2}), allowing a deeper physical understanding.
Let us begin, then, by giving the TBA equations in the
following form, in which the generalization of the already known
results for $[A_1]^{k,1}_-$~\cite{ItMo,Al3} and $[\cG]^{1,1}_{\pm}$~\cite{km}
clearly appears:
\eq
\nu_i^a(\theta) = \varepsilon_i^a(\theta) + \frac{1}{2\pi} \sum_{b=1}^r
\left[\phi^{ab}*L_i^b(\theta)-\sum_{j=1}^k (A_k)_{ij}
\psi^{ab}*L_j^b(\theta) \right]
\label{tba1}
\en
Here $i=1,...,k$, $a=1,...,r={\rm rank}\:\cG$, $L_i^a(\theta)$ is short
for $\log(1+e^{-\varepsilon_i^a(\theta)})$ and $A*B(\theta)$ denotes the
convolution $\int_{-\infty}^{+\infty}A(\theta-\theta')B(\theta')d\theta'$.
For the $[\cG]^{k,1}_-$ theories the energy
terms $\nu_i^a$ are chosen as $\nu_i^a=\delta_{i,1} M_a \cosh \theta$, where
$M_a = M\psi_a^{\cG}$ are the masses for the $k=1$ scattering theory.
$(\cG)$ stands
for the incidence matrix of $\cG$ and $\psi_a^{\cG}$ are the components of its
Perron-Frobenius eigenvector. The pseudoenergies $\varepsilon_i^a$
determined as solutions to this system, enter the formula for the free energy,
which in turn can be interpreted as the Casimir energy of the system on a
periodic strip
\eq
E(R) = \frac{\pi c(R)}{6R} =
-\sum_{a=1}^r \int_{-\infty}^{+\infty} L_i^a(\theta)
\nu_i^a(\theta)d\theta
\en
{}From this, the value of the central charge of the UV theory can be easily
extimated.
The two kernels $\phi^{ab}$ and $\psi^{ab}$ can be directly derived
from~\cite{BaRe} and
describe the scattering without color flip and with color flip respectively.
It is not surprising that $\phi^{ab}$ has the same form as for the
``colorless'' $k=1$ theories~\cite{km}
\eq
\phi^{ab}(\theta) = -i\frac{d}{d\theta} \log S^{ab}(\theta)
\en
where $S^{ab}$ is the diagonal S-matrix for $[\cG]^{1,1}_{\pm}$
theories, which
can always be put in the form~\cite{km,dorey}
\eq
S^{ab}(\theta) = \prod_{\alpha\in A_{ab}} f_{\alpha-\frac{1}{g}}(\theta)
f_{\alpha+\frac{1}{g}}(\theta) \virg f_{\alpha}(\theta)=\frac{\sinh\frac{1}{2}
(\theta+i\alpha\pi)}{\sinh\frac{1}{2}(\theta-i\alpha\pi)}
\en
where $A_{ab}$ is a set of rational numbers $\alpha$ with common denominator
$g$ (for details see~\cite{dorey}). If color
flip is allowed, then one has to take into account the second term too, where
a new kernel has to be introduced
\eq
\psi^{ab}(\theta) = -i\frac{d}{d\theta} \log T^{ab}(\theta)
\en
$T^{ab}$ is a fictitious S-matrix describing scattering with change of colors
$i\to i\pm 1$ (according to the admissibility $A_k$ Dynkin diagram appearing
in eq.(\ref{tba1})). Its general form is
\eq
T^{ab}(\theta) = \prod_{\alpha\in A_{ab}} f_{\alpha}(\theta)
\en
where the product runs over the {\em same} set $A_{ab}$ as for $S^{ab}$.
In~\cite{dorey2} it has been shown that $T^{ab}$ satisfies the same bootstrap
equations as $S^{ab}$. Introducing
\eq
P_{\alpha}(\theta)=-i\frac{d}{d\theta}\log f_{\alpha}(\theta)
=\frac{-\sin \pi \alpha}{\cosh \theta - \cos \pi \alpha}
\en
one can write
\eq
\phi^{ab}=\sum_{\alpha\in A_{ab}} (P_{\alpha-\frac{1}{g}}+
P_{\alpha+\frac{1}{g}}) \virg \psi^{ab}=\sum_{\alpha\in A_{ab}}P_{\alpha}
\en
Consider now the Fourier transform
\eq
\tilde{P}_{\alpha}(\kappa)=\int_{-\infty}^{+\infty}P_{\alpha}(\theta)
e^{i\kappa\theta}
d\theta = \left\{ \begin{array}{ll}
\frac{\sinh (1-\alpha)\pi \kappa}{\sinh \pi \kappa} & {\rm ~for~}\alpha >0 \\
0                                                   & {\rm ~for~}\alpha =0
\end{array} \right.
\en
By use of simple trigonometric identities connecting $\tilde{P}_{\alpha-
\frac{1}{g}}+\tilde{P}_{\alpha+\frac{1}{g}}$ to $\tilde{P}_{\alpha}$
one can prove the following identity
between Fourier transforms $\tilde{\phi}^{ab}(\kappa)$ and
$\tilde{\psi}^{ab}(\kappa)$
\eq
\frac{1}{2\pi}\tilde{\psi}^{ab}(\kappa) = \tilde R_g(\kappa)
\left[\delta^{ab}-\frac{1}{2\pi}\tilde{\phi}^{ab}(\kappa)\right]
\label{id1}
\en
where $\tilde R_g(\kappa)$ is the universal (i.e. indipendent of $a,b$)
function
\eq
\tilde R_g(\kappa)=\frac{1}{2\cosh \frac{\pi \kappa}{g}}
\en

The identity (\ref{id1}) together with another useful identity quoted
in~\cite{Al2}
\eq
\sum_{b=1}^r \left[ \delta^{ab} - \frac{1}{2\pi}\tilde{\phi}^{ab}\right]
\left[\delta^{bc}-(\cG)^{bc}\tilde{R}\right] = \delta^{ac}
\label{id2}
\en
allows to re-express system (\ref{tba1}) in a much more universal form.
Multiply the Fourier transform of both members of (\ref{tba1}) by
$\delta^{ad}-(\cG)^{ad}\tilde{R}$, then sum over $a$. Use of (\ref{id1}) and
(\ref{id2}) results in the following form for TBA equations
\eq
\nu_i^a = \varepsilon_i^a +\frac{1}{2\pi}\varphi_g*
\left\{\sum_{b=1}^r (\cG)^{ab}[\nu_i^b-\Lambda_i^b] -
\sum_{j=1}^k(A_k)_{ij}L_j^a
\right\}
\label{tba2}
\en
which is a direct generalization of the universal form found
in~\cite{Al2}. Here $\Lambda_i^a$ is short for $\log(1+e^{\varepsilon_i^a})$
and
\eq
\varphi_g(\theta) = 2\pi R_g(\theta) = \frac{g}{2\cosh\frac{g\theta}{2}}
\en
is exactly the same universal kernel of~\cite{Al2}.

By analogy with what done in~\cite{Al2,Al3,Al4,Al5}, it is interesting to
depict system (\ref{tba2}) in a graphical form. Here it is natural to consider
a ``bidimensional'' graph that can be thought as a ``product'' of the two
graphs encoded in $(\cG)$ and $(A_k)$ (see Fig.1). Non zero energy terms are
attached to the base ($i=1$) of the graph. The $k$ replicas of the
$(\cG)$ graph correspond to the $k$ different magnonic structures responsible
of exchange of colors. Colors can not change arbitrarily during scattering,
but only
according to the $(A_k)$ admissibility diagram. This graphical structure
helps to propose sets of TBA equations for $[\cG]^{k,1}_+$ too, along the
lines of ref.~\cite{Al4}. The simple modification is to take now $\nu_i^a =
\delta_{i,1}\frac{M_a}{2}e^{\theta}+\delta_{i,k}\frac{M_a}{2}
e^{-\theta}$. The physical interpretation, however is quite different:
the two pieces proportional to $e^{\theta}$ and to $e^{-\theta}$ are the
energy terms for massless left and right movers respectively. Again $M_a=
M\psi_a^{\cG}$, but the parameter
$M$ is here better interpreted as a crossover scale~\cite{ZamZam}.

Following then ref.~\cite{Al5} one can generalize even more the TBA to
include all the models $[\cG]^{k,l}_{\pm}$ just by taking energy terms as
follows
\eq
\nu_i^a(\theta) = \left\{ \begin{array}{ll}
\delta_{i,l}M_a \cosh \theta           & {\rm ~for~} [\cG]^{k,l}_- \\
\delta_{i,l}\frac{M_a}{2}e^{\theta} +
\delta_{i,k}\frac{M_a}{2}e^{-\theta}   & {\rm ~for~} [\cG]^{k,l}_+
\end{array}\right.
\label{nu}
\en
The graphs on which to encode the TBA are now $(\cG) \times (A_{k+l-1})$. The
most general TBA set of equation for the $[\cG]^{k,l}_{\pm}$ models then reads
as
\eq
\nu_i^a = \varepsilon_i^a +\frac{1}{2\pi}\varphi_g*
\left\{\sum_{b=1}^r (\cG)^{ab}[\nu_i^b-\Lambda_i^b] -
\sum_{j=1}^{k+l-1}(A_{k+l-1})_{ij}L_j^a\right\}
\label{tba3}
\en

To compute the free energy in the UV-limit, it is better to return to the form
(\ref{tba1}) of TBA eqautions, where of course now $\nu_i^a$ are given in the
more general form (\ref{nu}) and the sum over $j$ runs from 1 to $k+l-1$. The
UV behaviuor ($R\to 0$) of pseudoenergies is dominated by the solutions of
the shifted ``kink'' system~\cite{Al1,km,Al4}
\eq
\psi_a^{\cG}e^{\theta}\delta_{i,l} = \hat{\varepsilon}_i^a + \frac{1}{2\pi}
\sum_{b=1}^r \sum_{j=1}^{k+l-1}[\phi^{ab}\delta_{ij}-\psi^{ab}(A_{k+l-1})_{ij}]
*\hat{L}_j^k
\en
where $\hat{\varepsilon}_i^a$ are the shifted pseudoenergies such that
$\hat{\varepsilon}_i^a(0)=\varepsilon_i^a(-\infty)$.
This implies, after some standard manipulations, the following expression for
$c_{UV}$ in terms of Rogers Dilogarithm function
$\cL(x)=-\frac{1}{2} \int_0^x dt \left( \frac{\log(1-t)}{t} +
\frac{\log t}{1-t}\right)$
\eq
\begin{array}{lll}
c_{UV}={\displaystyle{\lim_{R\to 0}\frac{6RE(R)}{\pi}}} &=&
{\displaystyle{\frac{6}{\pi^2}\sum_{a=1}^r\left\{\sum_{i=1}^{k+l-1}\cL\left(
\frac{Y_i^a(-\infty)}{1+Y_i^a(-\infty)}\right) - \sum_{i=1}^{l-1} \cL\left(
\frac{Y_i^a(+\infty)}{1+Y_i^a(+\infty)}\right) \right.}}\\
&-& {\displaystyle{\left. \sum_{i=l+1}^{k+l-1}
\cL\left(\frac{Y_i^a(+\infty)}{1+Y_i^a(+\infty)}\right)\right\}}}
\end{array}
\label{dilog}
\en
where $Y_i^a(\theta)=e^{-\varepsilon_i^a(\theta)}$. $Y_i^a(\pm\infty)$ are
determined by the following algebraic nonlinear system of equations (use of
(\ref{id1}) and (\ref{id2}) for $\kappa=0$ has been invoked here)
\eq
y_i^a = \prod_{b=1}^r\left[(1+y_i^a)^{({\bf 1}-2K_{\cG})^{ab}}
\prod_{j=1}^Q (1+y_j^b)^
{(A_Q)_{ij}K_{\cG}^{ab}}\right]
\en
where $Q=k+l-1$ if $y_i^a=Y_i^a(-\infty)$ and $Q=l-1$ or $Q=k-1$ if
$y_i^a=Y_i^a(+\infty)$ in the second and third sum of eq.(\ref{dilog})
respectively, and $K_{\cG}$ is the inverse of the Cartan matrix of $\cG$.

The real solutions of this system satisfy Dilogarithm sum rules
\eq
\sum_{b=1}^r \sum_{i=1}^Q \cL\left(\frac{1}{1+y_i^a}\right) = \frac{rg(Q-1)}
{Q+g}
\en
Some of them are proven~\cite{BaRe}. The others can be checked
numerically with high precision. Together with the identity~\cite{lewin}
\eq
\cL\left(\frac{x}{1+x}\right) = \frac{\pi^2}{6}-\cL\left(\frac{1}{1+x}\right)
\en
they arrange the sum (\ref{dilog}) into
\eq
c_{UV}= c(\cG_k) + c(\cG_l) - c(\cG_{k+l})
\en
where $c(\cG_m)=m{\rm dim}\:\cG/(m+g)$ is the central charge of the
$\cG$-WZW model at level $m$, i.e. exactly what expected for the $[\cG]^{k,l}$
model from GKO construction~\cite{gko}.

The IR-limit is quite different for $[\cG]^{k,l}_-$ than for $[\cG]^{k,l}_+$.
In the first case the theory is dominated by massive kinks. One could
compute the free energy in the $R\to\infty$ limit and show that the expected
behaviour predicted by the interkink statistics is reproduced. We leave this
aspect for further work. We are more interested here to give evidence of
RG flows $[\cG]^{k,l}\to[\cG]^{k-l,l}$ described by the $[\cG]^{k,l}_+$ theory.
This simply follows by applying the same arguments as in~\cite{Al4,Al5}
(suitably generalized), to prove that
\eq
c_{IR}=c(\cG_{k-l})+c(\cG_l)-c(\cG_k)
\en
Notice that for $k=l$ this gives $c_{IR}=0$, thus signaling the fact that
$[\cG]^{k,k}_+ \equiv [\cG]^{k,k}_-$ is indeed massive (see ref.~\cite{Al5}).

Finally we would like to stress that the limit $k\to\infty$ at fixed $l$ of
$[\cG]^{k,l}_-$ TBA,
gives a TBA based on a semi-infinite tower of $\cG$ Dynkin diagrams. In analogy
to the comments in~\cite{Al5} this should correspond to the appropriate TBA
for the $\cG$ WZW model at level $l$ with asymptotically free perturbation
$J_a\bar{J}_a$. Indeed the UV central charge is $\frac{l {\rm dim}\:\cG}{l+g}$
in this limit. Of course such a limit can not be defined for the
$[\cG]^{k,l}_+$ TBA, while in this case it is reasonable to take the limit
of both $k,l\to\infty$, while keeping $k-l=n$ fixed. This, again generalizing
the ideas of~\cite{Al5}, gives an infinite (in both directions) tower of
$\cG$ Dynkin diagrams that encode a TBA suitable to describe the Principal
Chiral Model with Wess-Zumino term, which is known to give ${\rm dim}\:\cG$
free bosons at its UV limit, and the $\cG$ WZW model at level $n$ at its IR.
Indeed, the UV central charge is $c_{UV}={\rm dim}\:\cG$ here, while the IR
one amounts to $c_{IR}=\frac{n{\rm dim}\:\cG}{n+g}$. It would be interesting
to re-obtain these results by generalizing the approach of ref.~\cite{ZamZam}.

To conclude, making use of a diagonalization of the transfer matrix for IRF
models by Bazhanov and Reshetikhin, we have proposed here TBA equations for
all the $\cG_k\otimes \cG_l/\cG_{k+l}$ coset models perturbed by their
$\phi_{1,1,Adj}$ operator, both in massive and massless directions. We studied
the UV and IR limits, recovering the expected central charges. One should
also compute the next corrections to the UV limit of the free energy, in
order to estimate the bulk term (with or without logarithms), and to compare
the next regular terms with the informations from conformal perturbation
theory. Also, the corrections to the IR behaviour should be compared to the
(non renormalizable) perturbation expansion of the IR effective action for
massless theories and to the particle cluster expansion in the massive case.
We leave all this work and more details on the derivation of TBA to a
forthcoming more extensive paper.

\vskip .3cm
\noindent{\bf Acknowledgements and Notes added}

V.A.Fateev and Al.B.Zamolodchikov are also aware of the same
results~\cite{unp}, that they obtained in a very different way combining many
theories into a diagonal
S-matrix one, for which TBA can be easily obtained. The idea to study this
problem was generated during illuminating conversations with both of them.
For this and for many other discussions they are greatly acknowledged.

I am also very endebted to P.Dorey for many useful conversations and patient
explanation about properties of diagonal S-matrices, and to D.Bernard for
valuable discussions and for a careful reading of the manuscript.

While this work was towards its end in mid January 1992, a preprint by
M.Martins appeared~\cite{Mar},
where the particular case of $[A_2]^{k,1}_{\pm}$ is
studied in detail, and a TBA is proposed for $[A_N]^{k,1}_{\pm}$. Where they
overlap, his results are in perfect agreement with mine.

I am grateful to the Theory Group of C.E.A.-Saclay for the kind hospitality
they extended to me and to the Director of Sez. di Bologna of I.N.F.N., the
Theory Group of I.N.F.N.-Bologna and the Comm.IV of I.N.F.N. for the financial
support allowing me to spend this year in Saclay.

\newpage
\vskip 4cm
\setlength{\unitlength}{1mm}
\begin{center}
{\bf Fig.1 -} TBA graph for $[A_4]^{k,l}_{\pm}$ theory.\\
\vskip 2cm
\begin{picture}(65,110)(0,0)
\multiput(0,0)(10,0){4}{\multiput(0,0)(0,10){11}{\circle*{1}}}
\multiput(0,0)(10,0){4}{\line(0,1){20}}
\multiput(0,20)(10,0){4}{\multiput(0,0)(0,2){5}{\circle*{.3}}}
\multiput(0,0)(0,10){11}{\line(1,0){30}}
\multiput(0,30)(10,0){4}{\line(0,1){20}}
\multiput(0,50)(10,0){4}{\multiput(0,0)(0,2){5}{\circle*{.3}}}
\multiput(0,60)(10,0){4}{\line(0,1){20}}
\multiput(0,80)(10,0){4}{\multiput(0,0)(0,2){5}{\circle*{.3}}}
\multiput(0,90)(10,0){4}{\line(0,1){10}}
\put(-4,38){$l$} \put(35,38){$M_a \cosh \theta$}
\put(-4,68){$k$}
\put(42,108){$[A_4]^{k,l}_-$} \put(55,-6){$\nu_i^a$}
\put(60,38){$\frac{M_a}{2}e^{\theta}$}
\multiput(45,-2)(0,10){4}{0} \multiput(65,-2)(0,10){4}{0}
\multiput(45,48)(0,10){6}{0} \multiput(65,48)(0,10){2}{0}
\put(60,68){$\frac{M_a}{2}e^{-\theta}$}
\multiput(65,78)(0,10){3}{0} \put(62,108){$[A_4]^{k,l}_+$}
\put(12,-6){$A_4$}
\put(11,-9){$\longrightarrow$}
\put(-20,53){$\uparrow~A_{k+l-1}$}
\end{picture}
\end{center}

\begin{thebibliography}{99}
\bibitem{zam1}
A.B.Zamolodchikov, \IJMP{A4} (1989) 4235
\bibitem{zam2}
A.B.Zamolodchikov, Landau preprint (1989)
\bibitem{BeLec}
D.Bernard and A.LeClair, \NP{B340} (1990) 721
\bibitem{Al1}
Al.B.Zamolodchikov, \NP{B342} (1990) 695
\bibitem{Al3}
Al.B.Zamolodchikov, \NP{B358} (1991) 497
\bibitem{Al4}
Al.B.Zamolodchikov, \NP{B358} (1991) 524
\bibitem{Al5}
Al.B.Zamolodchikov, \NP{B366} (1991) 122
\bibitem{km}
M.Martins, \PL{B240} (1990) 485;\\
T.Klassen and E.Melzer, \NP{B338} (1990) 485; \NP{B350} (1991) 635
\bibitem{chr-rav}
P.Christe and F.Ravanini, \IJMP{A4} (1989) 897
\bibitem{FatAl}
V.A.Fateev and Al.B.Zamolodchikov, Paris preprint LPTHE 91-51
\bibitem{ItMo}
H.Itoyama and P.Moxhay, \PRL{65} (1990) 2102
\bibitem{Al2}
Al.B.Zamolodchikov, \PL{B253} (1991) 391
\bibitem{FatZam}
V.A.Fateev and A.B.Zamolodchikov, \IJMP{A5} (1990) 1025
\bibitem{DevFat}
H.J. de Vega and V.A.Fateev, \IJMP{A6} (1991) 3221
\bibitem{BaRe}
V.V.Bazhanov and N.Reshetikhin, J.Phys. {\bf A23} (1990) 1477;
Progr. Theor. Phys. Suppl. {\bf 102} (1990) 301
\bibitem{jmo}
M.Jimbo, T.Miwa and M.Okado, \MPL{B1} (1987) 73
\bibitem{ZamZam}
A.B.Zamolodchikov and Al.B.Zamolodchikov, Paris preprint ENS-LPS-355 (1991)
\bibitem{AlbMcc}
G.Albertini, S.Dasmahapatra and B.McCoy, Kyoto preprint RIMS-834 (1991)
\bibitem{dorey}
H.W.Braden, E.Corrigan, P.E.Dorey and R.Sasaki, \NP{B338} (1990) 689\\
P.Christe and G.Mussardo, \IJMP{A5} (1990) 4581
\bibitem{lewin}
L.Lewin, {\em Dilogarithms and associated functions}, Ed. MacDonald (London,
1958)
\bibitem{dorey2}
P.E.Dorey, \NP{B358} (1991) 654
\bibitem{gko}
P.Goddard, A.Kent and D.Olive, \PL{B152} (1985) 88
\bibitem{unp}
V.A.Fateev and Al.B.Zamolodchikov, to appear
\bibitem{AhBeLec}
C.Ahn, D.Bernard and A.LeClair, \NP{B346} (1990) 409
\bibitem{Mar}
M.J.Martins, Trieste preprint SISSA-91-EP-168
\bibitem{houches}
Lectures by: L.D.Faddeev - J.H.Lowenstein - H.B.Thacker, in {\em Advances in
Field Theory and Statistical Mechanics}, Proceedings of 1982 Les Houches
Summer School - J.B.Zuber and R.Stora Eds.
\end{thebibliography}
\end{document}